\documentclass[aps,pra,preprint,superscriptaddress]{revtex4}

\usepackage{graphicx}
\usepackage{amssymb}
\small
\begin{document}

\title{Casimir force calculations near the insulator-conductor transition in gold thin films}

\author{R. Esquivel-Sirvent}
\email[Corresponding author. Email:]{raul@fisica.unam.mx}
\affiliation{Instituto de F\'{\i}sica, Universidad Nacional Aut\'onoma
 de M\'exico, Apartado Postal 20-364, D.F. 01000,  M\'exico}

\date{\today}

\begin{abstract}

We present theoretical calculations of the Casimir force  for Au thin films near the insulator-conductor transition that has been observed experimentally. The dielectric function of the Au thin films is described by the Drude-Smith model. The parameters needed to model the dielectric function such as the relaxation time, plasma frequency and the backscattering constant depend on the thickness of the film.  The Casimir force decreases as the film thickness decreases until it reaches a minimum after which the force increases again. The minimum of the force coincides with the critical film  thickness where a percolation conductor-insulator occurs.

\end{abstract}

\maketitle

\section{Introduction\label{sec1}}

The attractive force between two parallel neutral plates made of a perfect conductor is known as the Casimir force. This force is explained in terms of the radiation pressure due to quantum vacuum fluctuations of the quantized electromagnetic field \cite{bordag} when boundaries are present.  Casimir's original derivation \cite{Cas48} substracted the vacuum energy between the plates, as obtained from the sum of allowed vacuum modes, from the vacuum energy when the plates were absent.  The resulting force per unit area is given by, 

\begin{equation}
F=-\frac{\hbar c \pi^2 }{240 L^4},
\label{cas}
\end{equation}
where $L$ is the separation between the plates.

In 1956 Lifshitz \cite{Lif56} generalized Casimir's results to real materials characterized by their dielectric function. In this theory,
the dissipative effects associated with the radiation reaction of the
elementary atomic dipoles composing the dielectric is balanced by the
fluctuating vacuum field in accordance with the
fluctuation-dissipation theorem.  If we consider two plates $i=1,2$ with different dielectric functions  the Casimir force is given by 

\begin{equation}
       \label{lifshitz}
       F=\frac{\hbar c }{2 \pi^{2}}\int_{0}^{\infty}Q dQ\int_{q>0}dk\frac{k^{2}}{q}(G^{s}+G^{p}),
\end{equation}
where $G_s= (r_{1s}^{-1} r_{2s}^{-1} \exp{(2  k L )}-1)^{-1}$ and $G_p=(r_{1p}^{-1} r_{2p}^{-1} \exp{(2  k L
)}-1)^{-1}$. In these expressions, the factors
  $r_{p,s}$  are the reflectivities for  $p$ or $s$ polarized light , $Q$ is the wavevector component along the
plates, $q=\omega/c$ and $k=\sqrt{q^2+Q^2}$.  

Experimentally the Casimir force has been measured by several groups 
 \cite{Lam97,Moh98,Ede00,Cha01,Bre02,Dec03,Dec05,Ian04} for separations  between the plates as low as $20$$nm$ \cite{george07}, using different techniques such as force balances, atomic force microscopes and micro mechanical balances. The experimental results verify the  Lifshitz theory to a high precision.  These experiments have also pointed out the importance that Casimir forces can have in micro and nano systems.  In particular the role it plays in stiction has been considered by several authors \cite{zhao,guo04,zhao07,esquivelapl,esquivelnjp,batra,delrio}.

The Lifshitz theory \cite{Lif56},  requires the knowledge of the dielectric function of the materials.  An important issue is  determining which  is the correct dielectric function that is consistent in describing the optical properties of the materials and the measurements of the Casimir force.  Although it may be thought that the problem is straight-forward, controversial results have been reported. The use of of the Drude model in Lifshitz theory has been argued to violate Nernst's heat theorem, while the plasma model presents no problem at all \cite{bezerra04}. This has been disputed by several authors . For example, spatial dispersion in the calculation of the Casimir force  \cite{sernelius05,hoye07} has been proposed to solve the problem or a non-vanishing damping constant in the Drude model at zero temperature . Recently,  an interesting proposal suggesting Johnson noise as a possible  finite-size effect to be included to solve this controversy was suggested for a system made of two parallel wires \cite{bimonte07}.   Also, Ellingsen \cite{ellingsen07} has established the criteria for the non violation of Nernst's heat theorem working on the real axis rather than the imaginary axis.   At this time the apparent violation of thermodynamics remains an interesting issue under discussion in the literature \cite{geyer08,foot}.

Several experiments and theoretical studies have been done to study how the different dielectric functions influence the value of the Casimir force. For example, an early attempt was made by Iannuzzi  that measured the Casimir force between hydrogen switchable mirrors (HSM) \cite{iannuzzi04}.  HSM's change their dielectric function dramatically when immersed in an hydrogen rich environment. However, this has been shown to have little effect on the Casimir force as the optical properties of the mirrors changed only in a narrow frequency range. These results were refined recently,  describing   the conditions needed to observe changes in the force between switchable mirrors \cite{man06}. Besides metals, semiconductors such as Si has been used \cite{irina07,mrssi} and light modulation of the Casimir force has been measured \cite{pinto99,inui, kika07}.  In this case,  light changes the carrier density of the semiconductor making it more or less metallic, thus changing the magnitude of the Casimir force.  
  Changing the magnitude of the force by a suitable choice of materials can be achieved using for example silicon based aerogels \cite{jap07}  that have the lowest index of refraction of any solid or with temperature changes of the dielectric function of materials such as   
 $VO_2$ that undergo a martensitic transition from dielectric to conductor. \cite{mrsvo,mohideensi,irina08}.    Recently, the possibility of using  metamaterials has also been considered. The effect of having a negative index of refraction is to have a repulsive Casimir force \cite{lambrecht08,yang08}.

Up to now, gold is the most common metal to work with in Casimir force experiments. When comparing with theoretical calculations, tabulated data for the dielectric function  is used \cite{palik}.  The films are usually thick enough to be able to use the assumption that the measured properties are close to those of bulk Au.   
At higher frequencies where quantum effects come into play, the dielectric function of Au can be described by an analytic model using a phenomenological approach that adds to the Drude model a Lorentz type response function \cite{etchegoin}. Another approach based on time dependent  density functional theory has been used to describe the intraband transitions of Au \cite{romaniello}.   Tabulated dielectric data for Au depends on the conditions of the sample preparation \cite{svetovoy06} and the Drude parameters obtained from the extrapolation to low frequencies gives different values for the plasma frequency and damping parameter. This implies variations of up to $5\%$ in the calculation of the Casimir force.

Another important parameter affecting the value of the Casimir force is the film thickness \cite{george07}.  This was shown experimentally by Iannuzzi \cite{lisanti05,lisanti06}  when it was demonstrated that the Casimir attraction between a metallic plate and a metal coated sphere depended on the thickness of the coating.  Most recently, 
  in reference \cite{svetovoy08}  different Au samples were prepared under similar conditions with thicknesses ranging from 120 nm to 400 nm.   From measured ellipsometry data it was verified that the plasma frequency varies from 6.8 eV to 8.4 eV for this set of particular samples, changing the Casimir force a few percent.  Also, the influence of slab thicknesses on the Casimir force was studied by Pirozhenko \cite{irina08} for slabs of finite thickness for several materials such as $VO_2$ and doped silicon   taking into account different carrier concentrations. 
 
 The reduction of size can significantly change  the physical parameters of a system such as the Debye temperature or the conductivity.   The case of clusters is well known. For example, a metallic cluster can go from being an insulator to a conductor depending on the size of the cluster \cite{jortner}.  A similar phenomenon is observed in metallic thin films.   As the thickness of the film approaches the mean free path,  the Debye temperature and the conductivity show a sharp decrease in its values.   This  was shown experimentally by Kastle \cite{kastle} with Au films whose thickness varied from 2 $nm$ to 70 $nm$.   Indeed, a conductor-insulator transition is observed as a function of film thickness in Au \cite{walther}.

In this paper, we study the Casimir between Au thin films near the insulator-conductor transition. The dielectric function for the Au films is described by the Drude-Smith model and the parameters for this model are obtained from reported experimental data.

\section{Dielectric function of Au thin films} 

 Thin films  grown by thermal deposition, present anomalous behaviors in  many of their physical properties, deviating from their bulk values with varying film thicknesses. In the case of the conductivity, 
 as the film thickness decreases it becomes less conductive until it reaches a point where it becomes insulating.   The reason for this change can be explained as a percolation transition. During deposition, disordered gold islands or clusters start forming on the substrate confining the conduction electrons to these islands.  As the  deposition continues, the fraction of island covering the substrate increases making it possible for electrons to hop between islands and increasing the conductivity. Finally,  more of these clusters are connected until a percolation threshold is reached \cite{walther}. After this percolation transition, the conductivity of the film will increase with increasing thickness until the bulk values are reached.  Experimentally this has been measured  by Walther \cite{walther} using terahertz time-domain spectroscopy.   
     
   An analytic expression of the dielectric function that describes the Au films near this transition is the Smith-Drude model.  This is a classical correction to the Drude model,  introduced by Smith \cite{smith} to describe the conductivity and dielectric properties of disordered metals, liquid metals and recently the metal-insulator transition of thin Au films \cite{walther}. 
This model assumes that the electrons in a metal are scattered with a  probability that follows a Poisson distribution. In this model, the current as a function of time after an impulse electric field is applied is ${\vec J}(t)={\vec J}(0)\exp{(-t/\tau)}f(n)$, where $f(n)$ is a function that has the information on the probability that an electron was scattered $n$ times in the time interval$[0,t]$ and $\tau$ is the relaxation time.  For the Drude-Smith model we have that
\begin{equation} 
f(n)=1+\sum_{n=1}^{\infty}\frac{c_n}{n!}\left(\frac{t}{\tau}\right).
\end{equation}
After a scattering event, the electron will retain only part of its initial velocity. This is represented by the value of the  coefficients $c_n$. The Drude model 
assumes that after each collision the electron has  a velocity not related with the velocity before the collision \cite{ashcroft} thus,  $c_n=0$ for all $n$.

 For a material with a plasma frequency $\omega_p$ and damping constant $\gamma=1/\tau$, the conductivity for the Drude-Smith model is 

\begin{equation}
\sigma(\omega)=\frac{\omega_p^2}{4\pi(\gamma-i\omega)}\left [ 1+ \sum_{n=1}^{\infty} \frac{\gamma ^n c_n}{(\gamma-i\omega)^n}      \right ]. 
\label{sigmasmith}
\end{equation}
The behavior predicted by the Drude-Smith model differs from that of the original Drude model  at low frequencies. Let $\sigma_D$ be the  conductivity predicted by the Drude model.  At zero frequency the DC current for the Drude-Smith model is $\sigma(0)=\sigma_D(0)(1+c1)$.  In the extreme case of $c1=-1$ the DC conductivity is zero.  

To calculate the Casimir force we need the dielectric function, which in the Drude-Smith case is given by
\begin{equation}
\epsilon(\omega)= 1-\frac{\omega^2_p}{\omega(\omega+i\gamma)}\left[ 1+\sum_{n=1}^{\infty}\frac{i^n \gamma^n c_n}{(\omega+i \gamma)^n}\right].
\label{epsilon}
\end{equation}

Typically only the term $n=1$ is enough to describe experimental results. If $c_1=0$ the Drude result is obtained  and the case where $c_1=-1$ corresponds to full backscattering of the carriers.  Experimentally, it has been observed that mercury is described with $c_1=-0.49$ and the value $c=-0.94$ corresponds to the quasicrystal $Al_{63.5}Cu_{24.5}Fe_{12}$ \cite{homes}.    

In the case of the Au thin films, that we consider in this work, we use Walther's \cite{walther} measurements  using teraHertz spectroscopy, that  show the crossing  from insulator to conductor as the film thickness increases.  The samples were prepared by thermal evaporation of Au and deposited on a Si substrate at room temperature and in vacuum. The conductivity measurements were best described by a Drude-Smith model. For films 20 nm thick the  bulk values  for the plasma frequency and damping parameter of Au  were recovered.  Decreasing the film thickness also reduces the plasma frequency until a critical thickness of $d_c=6.4 nm$ when the conductor-insulator transition is observed.  In table \ref{table1} we summarized the results of Walther \cite{walther} for some representative film thicknesses,  as well as the value of the constant $c$ used to describe the dielectric function. The damping parameter $\gamma$ does not change except close to the critical film thickness $d_c$.    

In Figure 1 we present the plot of  the Drude-Smith dielectric function   $\epsilon (i \omega)$ as a function of frequency for different Au film thicknesses.  The vertical axis is normalized to the function $ \epsilon(i\omega)_D$ that is the Drude dielectric function ($c=0$) for bulk Au with the parameters given by Walther \cite{walther}. We evaluated the dielectric function after  rotating the frequency axis to the complex plane $\omega \rightarrow i\omega$, that is used in the calculation of the Casimir force using Lifshitz formula \cite{Lif56}.  The three top curves in Fig. 1 show that as the thickness decreases the dielectric function decreases since, as seen from Table 1, the plasma frequency is also decreasing. However, at the critical thickness $d_c=6.4$ $nm$, the dielectric function is lower than the corresponding curve for the $d=4$ $nm$  curve in this frequency range. The lower curve (open circles) corresponds to the film thickness where the insulator-conductor transition occurs.  For lower frequencies,  we present in Figure 2 again the dielectric function for film thickness of $d=6.4$ $nm$and $d=4$$nm$. For the latter the dielectric function decreases and goes to zero meaning a zero DC conductivity since $c=-1$ for the thinnest of the films.   In both Figure 1 and 2 we have taken $\omega_0=10^{16}$ $s^{-1}$.  

Now we consider the effect of the film thickness on the Casimir force, between a system made of a half space of Au and a Au thin film deposited on a Si substrate.  The reduction factor $\eta$ defined from Eq. (1) and Eq. (2) as $\eta=\frac{F}{F_C}$ is a convenient way to see the behavior of the Casimir force.    To use Eq. (2) we calculate the reflection coefficients for a Au thin film of thickness $d$ on top of a substrate given by 
\begin{equation}
r_{p,s}=\frac{r^{01}_{ps}+r^{12}_{ps}e^{-2\delta}}{1+r^{01}_{ps}r^{12}_{ps}e^{-2\delta}}
\label{reflect}
\end{equation}
where $r^{i,j}_{p,s}$  are the Fresnel coefficients between material $i$ and $j$ where the subindex $0$ stands for vacuum, $1$ is the Au thin film and $2$ is the substrate.  The optical length is defined as \cite{born,irina08}
\begin{equation}
\delta=\frac{d}{c}\sqrt{\omega^2(\epsilon_1(i \omega)-1)+c^2 k^2}. 
\label{phase}
\end{equation} 

In Figure 3, we plot the reduction factor as a function of separation  using the parameters of the Au film from table 1. The dielectric function of the Si substrate is a Lorentz type oscillator model \cite{irina07}.   As expected, the Casimir force decreases with decreasing  film thickness.  However,  this trend changes after the critical film thickness of $d=6.4$ $nm$ where the percolation transition occurs.  At this critical distance the plasma frequency attains a minimum value, as seen in table 1.  After  the critical thickness the force increases again although the film thickness is decreasing. The behavior at the distance $d_c$ where we see a sudden increase in the relaxation time and a decrease of the plasma frequency is consistent with the singular behavior in the dielectric constant in metal-dielectric percolation transitions \cite{drubrov,grannan}.  This behavior is seen more clearly  in Figure 3, where we show the reduction factor as a function of film thicknesses when keeping the separation between the plates fixed at  $L=400$ $nm$.    
   
 Following  previous works by  Piroshenko and Lambrecht \cite{irina07,irina08}, we further analyze the effect of the thin film  using the optical length  or phase factor $\delta$ of the Au thin film \ref{phase}.  In particular, the difference between using the Drude, Plasma or Drude-Smith model can  be seen in the behavior of the optical length with frequency.  For example at the critical distance $d_c$ we calculate the optical length assuming a simple plasma model $\delta_p$, the Drude model $\delta_d$ and for the Drude-Smith model,  and calculate the percent difference $\Delta=100\times|\delta-\delta_p|/\delta$ and $\Delta=100\times|\delta-\delta_D|/\delta$. This is shown in Figure 5, where we plotted the percent difference as a function of frequency. As expected, for high enough frequencies the plasma model is enough to describe the dielectric function of the Au thin film, if interband transitions are ignored. However,  at  low frequencies the  percent difference between the phase factor when the Drude or plasma model are used becomes larger in particular for the plasma model.   The optical length was calculated for the particular  value $ck/\omega_0=1$ as in reference (\cite{irina08}).
 
 \section{Conclusions}
 
 In this work we studied the effect that the film thickness has on the Casimir force for films near the critical thickness where a conductor-insulator transition occurs.  To describe the dielectric function of the Au thin film we use the Drude-Smith model  that describes the experimental data available in the literature.  The parameters used in our calculations are best-fit parameters from experimental measurements. For thick films the force decreases with decreasing film thickness until a critical thickness is reached after which the Casimir force increases even with decreasing film thickness.     The  minimum value of the Casimir force is due to a decrease in the plasma frequency and an increase in the relaxation time when the metal-insulator transition occurs in the film. The use of Au thin films has the advantage that the growing techniques are well known, and experimentalist working on Casimir force measurements use gold coated spheres and substrates routinely. Depending on   the film thickness the dielectric function deviates from the typical Drude and plasma models in accordance to the Drude-Smith model.  As the thickness decreases the value of the backscattering constant $c$ decreases changing the value of the DC conductivity. This could lead to  experimental settings that further explore the role of DC conductivities and finite temperature controversies.

 \acknowledgements{ The author wish to acknowledge the helpful comments of L. Reyes-Galindo.   Partial support for this work comes from  DGAPA-UNAM project 113208.}

  \newpage
  \begin{table}[h]
  \begin{ruledtabular}
	\begin{tabular}{lllll} 
	\hline
		d(nm) & $\omega_p \times 10^{15}$$ s^{-1}$ & $1/\gamma$ (fs) & c&$\sigma/\sigma_D $\\ 
		\hline
		20 & 13.19 & 18 & 0 & 1\\ 
		15 & 10.05 & 19 & 0 &1\\ 
		10 & 6.28 & 19 & 0 &1\\ 
		6.4 & 1.25 & 80 & -0.7 &0.3\\ 
		4 & 1.88 & 20 & -1&0\\
		\hline
	\end{tabular}
	\end{ruledtabular}
	\caption{Best fit parameters for the Drude-Smith model taken from Ref. (\cite{walther}). The last column shows the ratio of the DC conductivity for the Drude-Smith model to the expected DC Drude model conductivity.}
	\label{table1}
\end{table}

\newpage

  \begin{figure}[h]
\includegraphics[width=.8\textwidth]{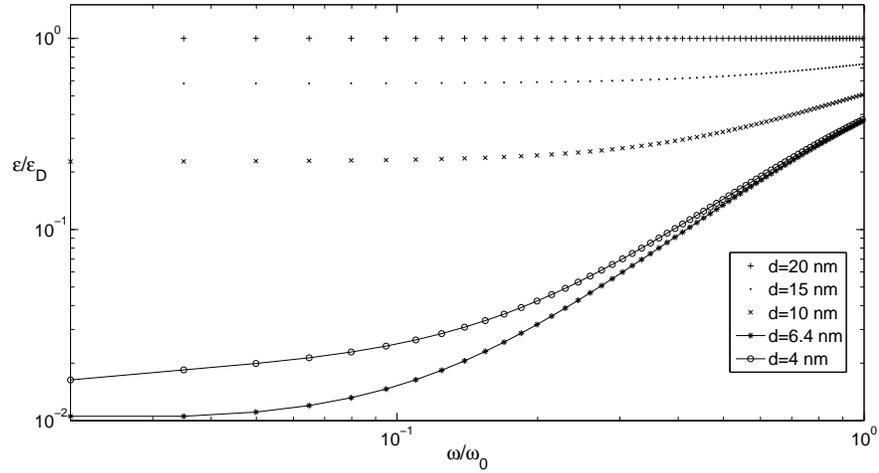}

\caption{\label{fig1} Dielectric function for thin films using the Drude-Smith model as a function of frequency. The parameters for each film thickness were taken from experimental data. The dielectric function is compared to the bulk Drude dielectric function reported in the experimental data of Walther \cite{walther}. In all figures we take $\omega_0=10^{16}$$1/s$. }  
\end{figure}

 \begin{figure}[h]
\includegraphics[width=.8\textwidth]{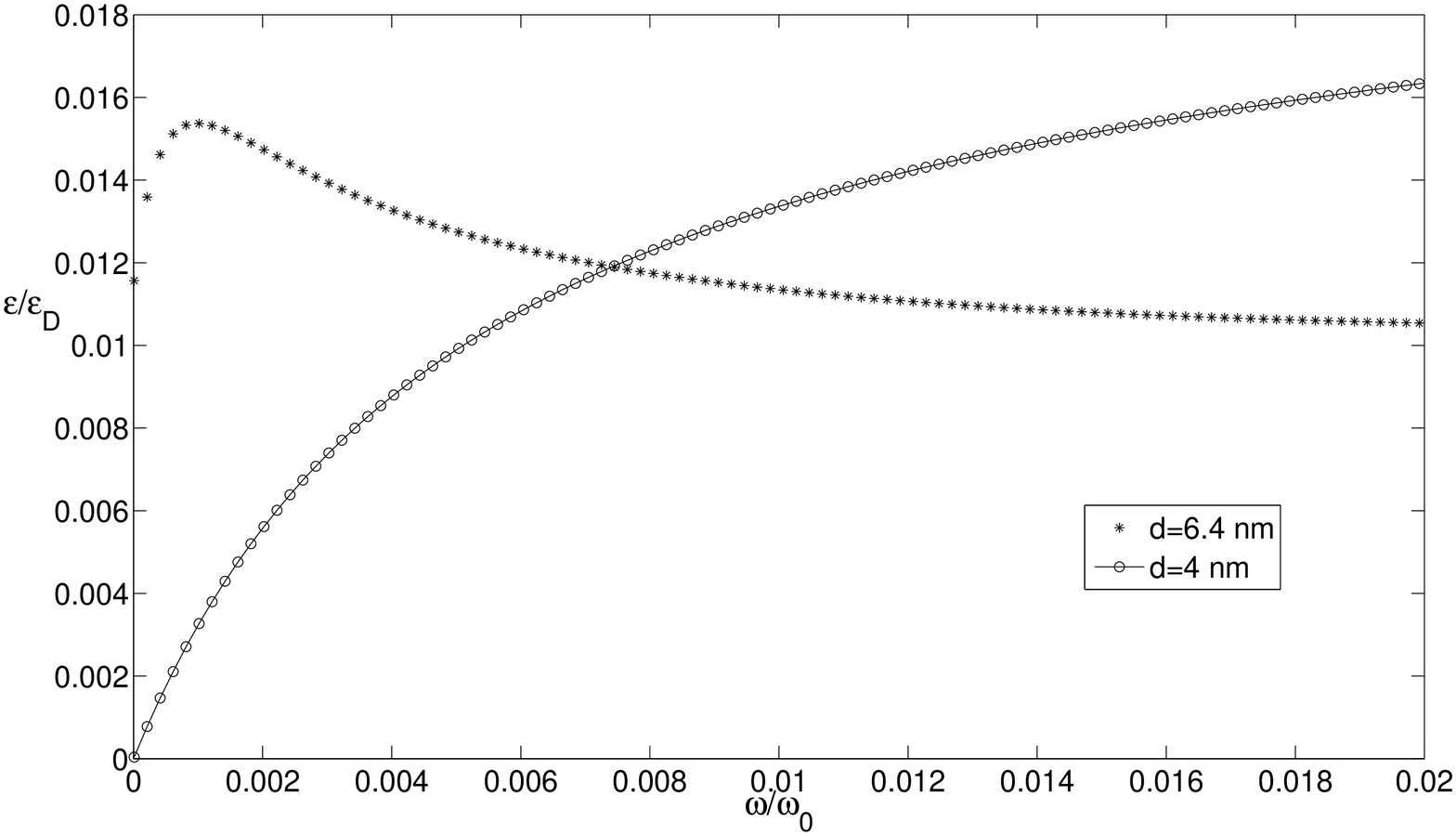}

\caption{\label{fig2} In this figure we plot again the dielectric function for low frequencies of the two lower curves of Fig.1. This is for $d=6.4$ $nm$ and $d=4$$nm$. The dielectric function for the thinner film decreases We see a cross over since for $d=4$$nm$ the backscattering constant is $c=-1$ implying a zero DC conductivity and hence $\epsilon(i\omega)\rightarrow0$. }  
\end{figure}

 \begin{figure}[h]
\includegraphics[width=.8\textwidth]{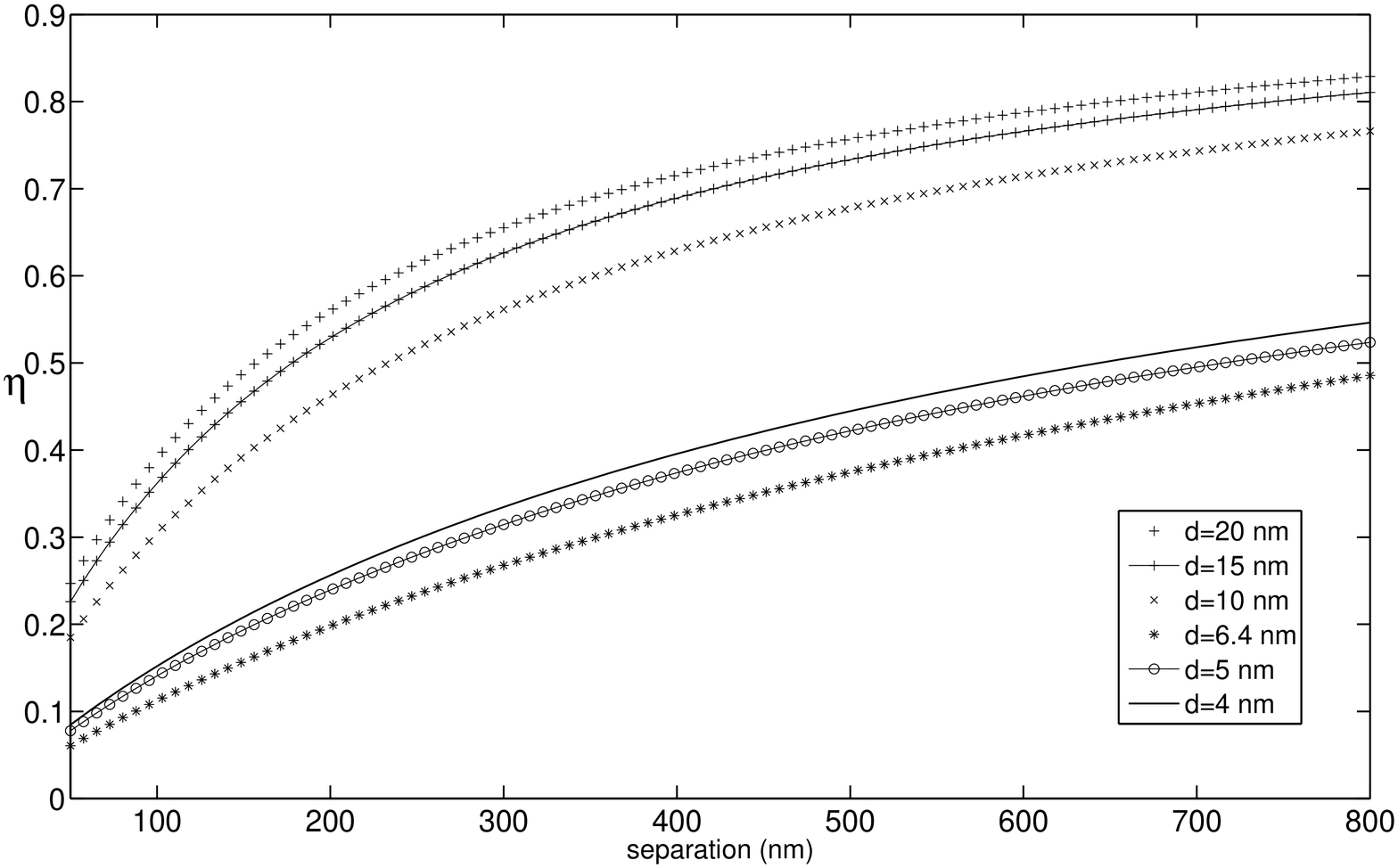}

\caption{\label{fig3} Reduction factor as a function of frequency for different  Au film  thicknesses. As the film thickness decreases the Casimir force also decreases until it reaches the critical thickness $d=6.4 nm$, after which the force increases even with decreasing film thickness. }  
\end{figure}

 \begin{figure}[h]
\includegraphics[width=.8\textwidth]{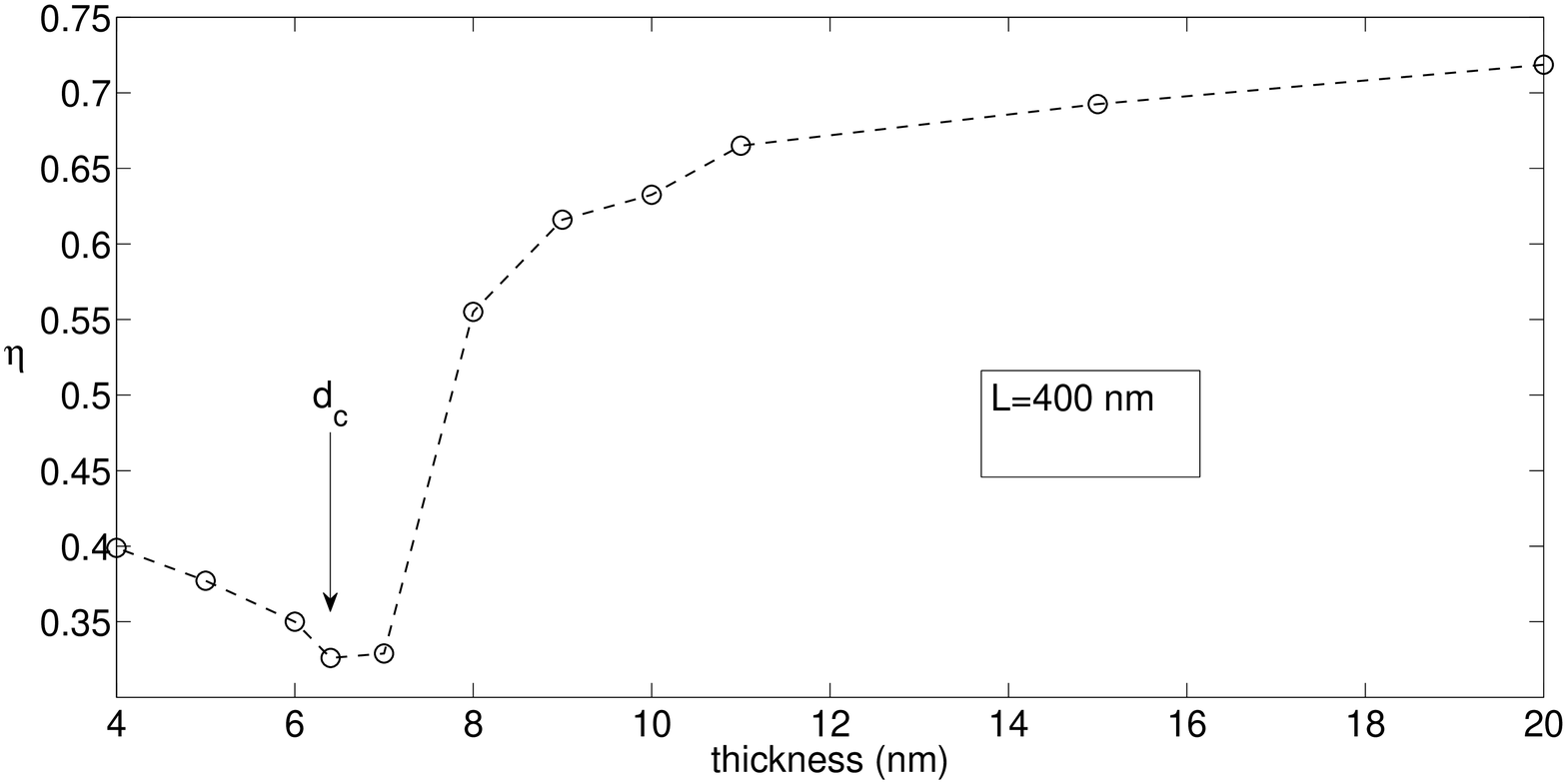}

\caption{\label{fig4} Reduction factor as a function of the Au film thickness. The reduction factor is calculated assuming a separation between the plates of $L=400$$nm$ and using the optical data for the Au films from Ref.( \cite{walther}).  At the thickness of $d_c$ the percolation transition occurs, and the Casimir force attains a minimum. }  
\end{figure}

\begin{figure}[h]
\includegraphics[width=.8\textwidth]{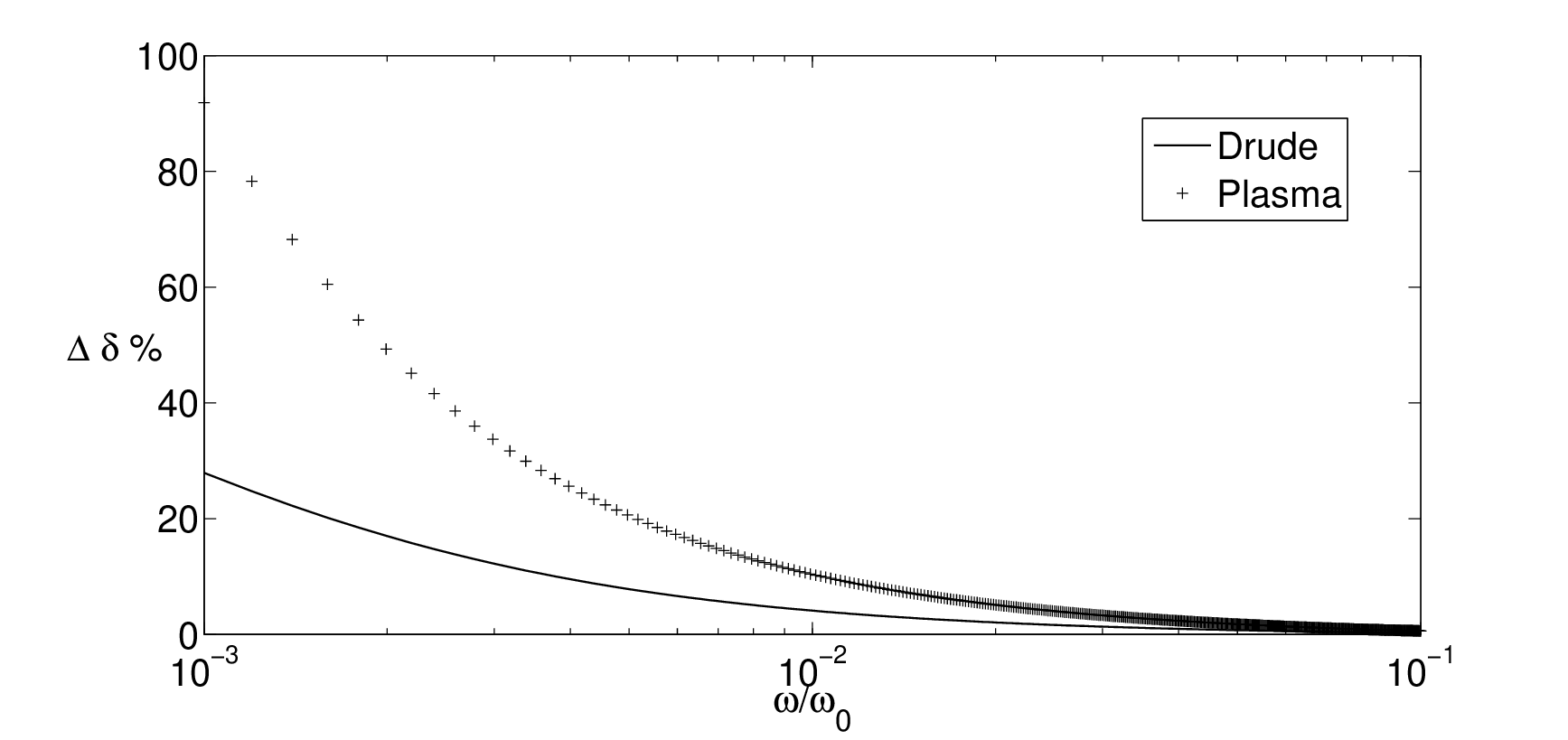}

\caption{\label{fig5}  Percent difference  between the optical length for the Drude-Smith model and the Drude model (solid line ) and the plasma model (crosses).  The curves are calculated for a Au film of thickness $d_c$. The low frequency behavior shows strong  differences if the plasma or Drude model are used. In this figure  $\omega_0=10^{16}1/s$ and  $ck/\omega_0=1$.  }  
\end{figure}


\begin{thebibliography}{99}
\bibitem{foot} The number of papers, comments and replies on the problem of Nernst heat theorem and the Casimir force in the past five years shows that this problem is of importance to the Casimir community. The references that have been cited in this paper are only a small sampler of representative works.
\end{thebibliography}
\end{document}